\documentclass[10pt]{article}
\usepackage{graphicx}
\begin{document}

\begin{center}
\Large{\bf Periodicities of Quasar Redshifts in Large Area Surveys}
\end{center}

\begin{center}
{\bf H. Arp}\\
Max-Planck-Institut f\"ur Astrophysik, 85741 Garching, Germany
\end{center}

\begin{center}
{\bf D. Roscoe}\\
Applied Maths Dept, Sheffield University, Sheffield S37RH, UK
\end{center}

\begin{center}
{\bf C. Fulton}\\
Centre for Astronomy, James Cook University, Townsville Queensland 4811, 
Australia.
\end{center}

\bigskip

\date{Received; 2004}

\begin{abstract}

We test the periodicity of quasar redshifts in the  2dF and SDSS surveys. In the overall surveys
redshift peaks are already apparent in the brighter quasars. But by analyzing sample areas in 
detail it is shown that the redshifts fit very exactly the long standing Karlssson formula 
and confirm the existence of preferred values in the distribution of quasar redshifts.

We introduce a powerful new test for groups of quasars of differing redshifts 
which not only demonstrates the periodicity of the redshifts, but also their physical 
association with a parent galaxy. Further such analyses of the large area surveys 
should produce more information on the properties of the periodicity.

\end{abstract}

{\it Subject headings} galaxies: active -- quasars: general

\section{Introduction}

Hawkins, Maddox and Merrifield (2002) claimed that an analysis of
the large 2dF quasar sample showed no evidence for quasar redshift
periodicity. The purpose of the present paper is not to address their
criticism of past evidence (that has been done by Napier and Burbidge,
2003, who show the past evidence is indeed valid) - but 
rather to show that, also contrary to the Hawkins et al.
claim, a simple qualitative analysis of the new data plus a detailed
analysis of a few sample fields, reveals that quasar periodicity is
indeed strongly present.

\subsection{The Basic Hypothesis}

The model which is used here to predict properties of quasars in the survey regions
is radically different from the canonical view. It has evolved from observational 
evidence over the past thirty eight years and has two main properties which can be 
expressed as: \begin{itemize}

\item Quasars originate from within, and are ejected by, active galaxies. Thus,
in a real sense, active galaxies are parents to quasars; 

     The evidence for this arose originally from radio quasars paired across disturbed
galaxies. It was acccepted that galaxies ejected radio sources 
in roughly opposite directions. Ejection therefore appeared a natural explanation for 
these pairs of radio quasars (Arp 1967). Later X-ray jets were often found in the cores 
of radio jets and they were in turn associated with the more numerous X-ray quasars 
in pairs and lines - many of which were found to be aligned with the nuclei of low 
redshift galaxies.

 Some of the papers generally showing evidence involving radio emitting QSOs
and bright galaxies, X-ray-emitting QSOs and active galaxies, and
pairs with optical, radio and X-ray connections are (Burbidge et al. 1971; 
Pietsch et al. 1994; Burbidge 1995, 1997, 1999; Radecke 1997; Arp 1967, 
1996, 1997,1999, 2003; Arp et al. 1990, 2002). Two of the most impressive recent 
examples are the X-ray QSOs connected to the nucleus of NGC 3628 (Arp 
et al. 2002) and the discovery of two QSOs in the optical bridge between NGC 7603
and its companion galaxy (L\'{o}pez-Corredoira and Guti\'{e}rrez 2002).

The observed physical association with low redshift galaxies is explained by the 
ejection hypothesis but it also enables an empirical test of the periodic nature of the 
quasar redshifts because;

\item 
Quasars are always higher redshift than the ejecting galaxy and none show negative
redshifts of approaching velocities. The conclusion is that they must have intrinsic 
redshifts which are much larger than their ejection velocities.  Supporting the
interpretation of the redshifts as intrinsic it was then found that the redshifts had 
discrete preferred values in the frame of their parent galaxy. (The Karlsson
periodicity formula).The ejection velocities 
allowable are small compared to these intrinsic redshifts. They are observed to 
spread the redshifts around the preferred values by less than about +/- 0.1 in z.
from their preferred values. Nevertheless that deviation from the periodic value can be
used to test the prediction that the velocities are generally + and - in pairs,
representing the approaching and receding velocities which would be necessary in
order to conserve momentum in an ejection process. We test that prediction in the
present paper by analyzing some cases in which several pairs of quasars surround 
a central, ejecting galaxy.

\end{itemize} Three specific consequences follow:

\begin{itemize}
\item Firstly, since
quasars are (by hypothesis) associated with {\it specific} active galaxies, then
any redshift periodicity will only become apparent when the quasar's redshift is
corrected into the rest-frame of the parent galaxy.
\item Secondly, the putative parent galaxies will be generally much
brighter than their quasar off-spring;
\item Thirdly, quasars of bright apparent magnitude will
generally be nearby so that their associated parent galaxies  {\it will tend} to
be low-redshift objects. Quasars of faint apparent magnitude will generally be
more distant so that their associated parent galaxies {\it will tend} to be
higher-redshift objects. This implies that, {\it generally speaking}, the
redshifts of quasars with bright apparent magnitude will require little or no
correction for the periodicity effects to be manifested, whilst quasars with faint
apparent magnitude will require considerable correction.
\end{itemize}

We begin by reviewing the history of redshift periodicity
claims, and then consider the recent counter-claim by Hawkins, Maddox and
Merrifield (2002) that, based on a large-scale analysis of the 2dF field, quasar
redshift periodicity is not present. For this we first consider the overall 2dF data, 
pointing out that periodicity is apparent in the bulk data and then go on to 
consider, in detail, four sample fields of 2dF and SDSS.  We conclude that the 
basic hypothesis is very much strengthened by the new large-scale surveys.

\subsection{A Brief History} The existence of a preferred
redshift of $z = 1.95$ for bright quasars was first pointed out by Burbidge and
Burbidge (1967). Further periodicities were found by Lake and Roeder (1972) and
Burbidge and O'Dell (1972). Karlsson (1971, 1973, 1977) showed that the formula
$\Delta \log_{10}(1 + z) = .089$  provides a good fit to the major observed
peaks at $$z = 0.30, 0.60, 0.96, 1.41, 1.96, 2.64 . . .$$

Karlsson established his fit with a sample of 574 quasars to a 99.9\% confidence level.
Later Barnothy and Barnothy (1976) confirmed this periodicity to the 99.99\% 
confidence level. Depaquit et al. (1985) showed how every test to that date supported 
the Karlsson formula. With a larger sample of quasars Fang et al. (1987) again 
confirmed these results. A few years later it was shown that the periodicity was
particularly prominent for quasars which were close (by angular measure) to low
redshift galaxies (Karlsson 1990; Arp et al. 1990).

It had also become clear by then that those low redshift galaxies with which
such quasars appeared to be preferentially associated, tended to be
morphologically disturbed and/or to have spectroscopically active nuclei. For
example, the pairing of X-ray quasars across X-ray active Seyert galaxies
demonstrated physical association at the 7.5 sigma level (Radecke 1997; Arp
1997).

Evidence continued to accumulate, culminating in the discovery of two active Seyfert 
galaxies, $NGC \,3516$ (Chu et al. 1999) and $NGC\, 5985$ (Arp 1999) which had a total 
of eleven quasars accurately aligned along their minor axes. Ten of these clearly
associated quasars fell close to the Karlsson peaks (see Arp 1998, p. 244, 284
-285). Most recently Burbidge and Napier (2001) have tested the periodicity yet
again with three different samples of quasars and found the fit to be
significant at the $10^{-5}$ level.

The earlier studies, such as those discussed above, used largely bright quasars
(and therefore nearby, according to the hypothesis) so that the associated
(putative parent) galaxies had negligible redshifts. It follows that, for
these samples, the periodicities (if they exist) would show up in the {\it
uncorrected} redshifts. It is important to recognize that as early as 1990 it
was understood that quasar redshift periodicity appeared to decline as
quasar apparent brightness declined. This result became clear in the process of
addressing the question of color selection effects in catalogued quasars.

Two different samples of quasars were tested by Arp, Bi, Chu and Zhu (1990).
In the first sample, multiple quasars falling
significantly closer than chance to low redshift galaxies were analyzed. The
discovery criterion for these quasars was blue $U - B$ colours which were shown
to be insensitive to redshift values. The significance of the peaks ranged from
94\% to 99.5\% for a sample size of 54 to 49 quasars. It was found, however,
that the redshifts showed better correspondence to the Karlsson peaks when
corrected for the red shift of the associated galaxy. The effect was small,
however, because the redshifts of the associated galaxies were also
relatively small.

The redshift transformation formula is the canonical one
$$(1 +z_0) = (1 + z_Q)/(1 + z_G),$$ where $z_Q$ is the observed quasar redshift
and $z_G$ is the observed redshift of the associated galaxy which, by our
hypothesis, is the ejecting galaxy.

The second sample analysed consisted of all radio quasars brighter than 1
Jansky at 11cm. It turned out in the end that this sample gave a fit to the
Karlsson formula at a level of 99.997\%. Of course, for quasars chosen on the
basis of their strong radio emission there could be no question of color
selection. But one thing was clearly stated in the Arp et al. 1990 paper and
that was the redshifts {\it". . . become less clearly periodic as the radio
strength declines. . ."}  Also, it was stated that: {\it "Likewise the
periodicity becomes less well marked fainter than V $\sim$ 18 mag. . ."}

With both the radio and optical apparent brightness diminishing it was
reasonable to suppose that the quasars were becoming more distant. The
parent galaxies with which they were associated would then be more distant and
their redshifts would become non-negligible. The fainter quasars would need
appreciable corrections to bring them into the rest frames of their parent
galaxies. We would then expect any plot of quasar redshifts against apparent
magnitude to show the periodicity better at bright magnitudes than at faint
magnitudes. As we shall show, this is {\it exactly} what we find with the 2dF
quasars.

\subsection{Hawkins et al Analysis of 2dF QSOs} After all this, a sample
of 1647 quasars close, by angular measure, to galaxies of otherwise arbitrary
properties in the 2dF survey field of 289.6 sq. deg.  was announced as showing
no periodicity "at the predicted frequency. . .  or at any other
frequency."(Hawkins, Maddox and Merrifield 2002).

We note that the sole criterion employed by Hawkins et al to identify a
parent galaxy was simply projected closeness on the sky of the galaxy to
the quasars tested - thus, apart from the fact that, for the host of faint galaxies, there 
are very many more purely chance associations than for bright galaxies, critical 
criteria concerning the relative brightness and the state of activity of the 
chosen galaxies were not considered. In fact present investigations show 
the most significant associations of faint survey quasars is, for example, with UM 
strong emission line galaxies(Univ. Michigan objective prism survey). These 
parent galaxies are are in a brightness range around 15 to 18 mag. - not in 
the brightnes range of the 2dF galaxies in which Hawkins et al. were attempting 
to test correlations.

\section{The $2dF$ Fields}
We show  that  quasar density plots in the apparent magnitude/redshift plane
reveals the periodicity structure crudely but plainly, and in a way that
requires no sophisticated statistical analysis to appreciate.

In essence, we proceeded as follows: The 2dF survey was conducted in two
distinct declination strips: the declination strip $-2.5^o \leq Dec. \leq +2.5^o$ centered at
about $RA=12h15m$ contained in the first release 4200+ quasars. The declination strip
$-32^o \leq Dec \leq -27^o$ centered at about $00h30min$ contained 6700+
quasars.

\begin{figure}
\includegraphics[  width=3.5in,
  height=3.5in]{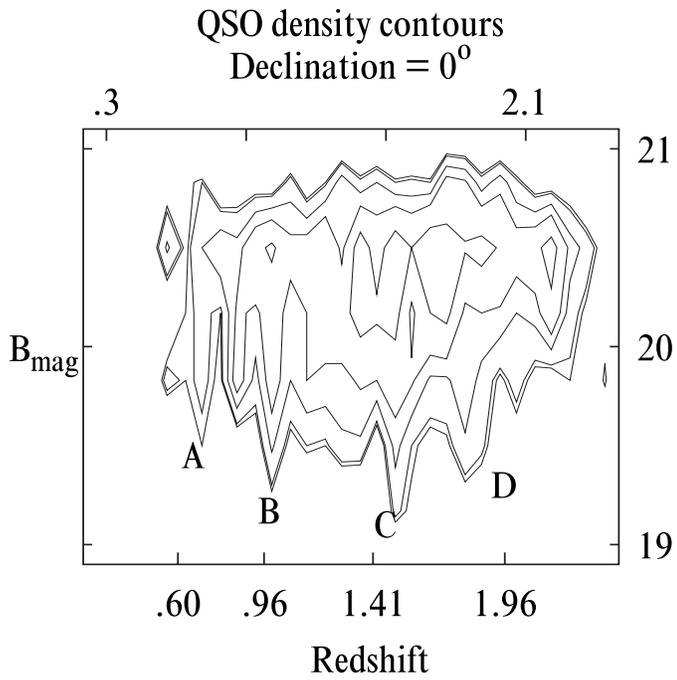}
 \caption{Apparent magnitude vs measured redshift plot for quasars in the 2dF
survey declination strip  centered on Dec = 0 deg. The contour lines
represent quasar density where the outermost contours represent low densities,
and the innermost contours represent high density.
Note that  peaks are
only present for the brighter quasars.} \label{Fig1}
\end{figure}

\begin{figure}
\includegraphics[  width=3.5in,   height=3.5in]{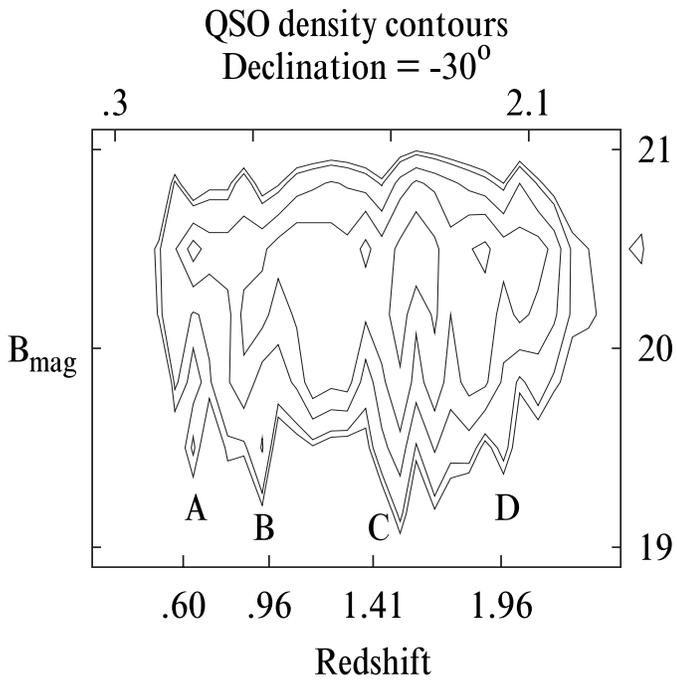}
 \caption{Apparent magnitude vs measured redshift plot for quasars in the 2dF
survey declination strip  centered on Dec = -30 deg.    The contour lines
represent quasar density where the outermost contours represent low densities,
and the innermost contours represent high density.
The expected Karlsson peaks at $0.60, 0.96, 1.41, 1.96$  are indicated }
\label{Fig2}  \end{figure}

For each declination strip, the whole region was divided into boxes $\Delta z
\times \Delta B = 0.075 \times 0.3$ in the redshift/apparent magnitude plane,
and the number of quasars per box
counted. Figures \ref{Fig1} and \ref{Fig2} show density contours plotted through
the fields. Since we are using the {\it raw} redshift
data then, according to the basic
hypothesis, at any given concentration level we should then expect to see
contour spikes {\it for the brighter objects} near the Karlsson peaks $z=0.3, 0.6,
0.96, 1.41,1.96, ...$ whilst, for  dimmer objects we should expect to see no
signal at all.  These expectations are met, as we describe below.

\begin{figure}
\includegraphics[  width=3.5in,   height=3.5in]{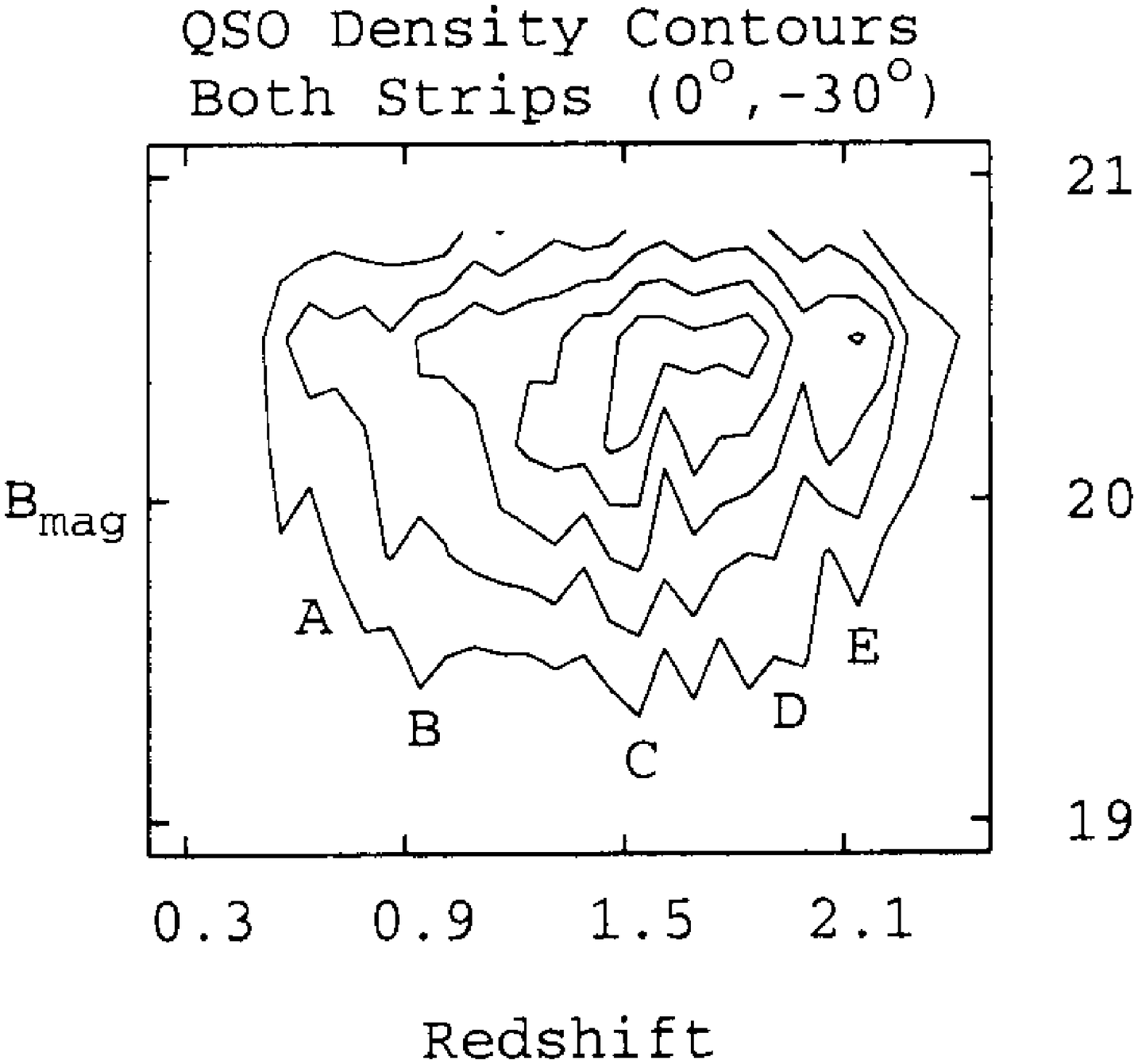}
 \caption{Apparent magnitude vs measured redshift plot for quasars in the 2dF
final release (22,435 QSO's). The canonical Karlsson peaks at $0.60, 0.96, 
1.41, 1.96$  are indicated by their letter positions in Figs 1 and 2. An additional
peak at $z \sim 2.1$ is also indicated, as noted in text}
\label{Fig2a}  \end{figure}

The  declination strip $-2.5^o \leq Dec. \leq +2.5^o$ is plotted in
Figure \ref{Fig1}, and the outermost contour represents a density of 20 quasars
per box whilst the innermost contour represents a density of 50 quasars per box.
In this region (in the direction of the Local Super Cluster) we see strong
bright-end spikes near $ z = 0.6, 0.9, 1.5$ and $1.9$. These are labelled
$A,\,B,\,C$ and $D$ respectively. 

The  declination strip $-32^o \leq Dec. \leq -28^o$ is plotted in Figure
\ref{Fig2}, and the outermost contour represents a density of 27 quasars
per box whilst the innermost contour represents a density of 65 quasars per
box. In this region  we see again strong bright-end spikes near  $ z = 0.6,
0.9, 1.5$. The situation near $z=1.9$ is less clear. These are also labelled
$A,\,B,\,C$ and $D$ respectively. 

The really striking feature, of course, is the consistency of positioning
between the $A,B,C$ bright-end spikes in both figures - it is this feature which
makes it clear that these spikes are not functions of random variation in the
data, but represent a real phenomenon. Together, they provide a strong,
if noisy, confirmation of the reality of the Karlsson peak in quasar redshift
data in the large sample presented in the first release of the 2dF .

\subsection{The Final 2dF Release}

After completion of this paper the final release of the 2dF fields took place 
(Colless et al 2003). It was thought interesting to see the foregoing analysis
applied to the final data set. This is supplied here in Fig. 3 where all 22,435 QSO's, in both 
the $Dec. = 0^o$ and $Dec. = -30^o$ strips together, are contoured in density steps 
of 102, 135, 165, 195 and 225. Again we see the peaks B,C and D roughly confirming 
the early release results.

It is clear, however, that in different regions different bright spikes are encountered
and that with averaging over many regions they tend to become less distinct. For example
in Fig. 3 there is a new prominent spike at a little greater than z = 2.1 labeled E. This
can be identified with the quasars from a large region near the south galactic pole
which is dominated by the Sculptor Group of galaxies.  Between roughly 
$23h\leq R.A. \leq 01h$  and $-20d\geq Dec. \geq-40d.$, objective prism surveys in
strips across this region showed concentrations of $z\geq 2$ quasars. (Osmer and Smith
1980; Arp 1983; 1984). The current 2dF strip across this region confirms this previously 
discussed concentration of quasars.

However, it is also clear from these contour plots that even selecting a bright
subsample that was sufficiently large to include all four bright-end spikes
(typically, $B_{mag} < 19.6$) {\it would not} produce a sample that was
sufficiently clean that a simple technique like power spectrum analysis 
would detect the periodicity signal implied by the $A,B,C,D$ spikes. For example
inclusion of somewhat fainter, uncorrected quasars moved the predicted peaks 
in Figs. 1 and 2 toward z = .65, 1.02, 1.48 and 2.05, (as predicted by the results of  
Arp et al. 1990, especially Fig. 3a and 3b of that paper). For much fainter apparent 
magnitude quasars the parent galaxies would have a much wider range in redshift and 
the raw quasar redshifts would be expected to approach a smooth distribution as is, 
in fact, born out by the contours at faintest apparent magnitudes in Figs. 1, 2 and 3.

It would seem that,  for the basic hypothesis to be comprehensively
demonstrated, it is necessary to attempt the association of quasars with their
respective {\lq parent'} galaxies, so that the quasar redshifts can be
transformed into the appropriate rest frames. For this purpose, there would seem
to be no substitute for actually looking at the fields which are being tested to
see what kinds of galaxies there are in those fields, and how the quasars are
actually distributed with respect to those galaxies. 

\section[]{A Test in a Field of the Sloan Digital Sky Survey}
In the following, we use a single field from the $SDSS$ (Abazajian et al. 2003) 
to make the point that,whereas a naive analysis would find no evidence of 
periodicity, a more critical analysis which paid full attention to the basic 
hypothesis would find it to be very strongly confirmed {\it on this single field}.

The field to be considered, which is in the vicinity of the active ($z=0.017$)
Markarian galaxy $NGC\, 622$, was chosen for two reasons:
\begin{itemize}
\item firstly, prior to the $SDSS$, there were known to be  Karlsson-peak
quasars at $z = 0.91$ and $1.46$ only $71''$ and $73''$ away (Arp 1981; 1987).
\item  secondly, in the intervening years, the $SDSS$ survey has identified
further quasars in the general vicinity of $NGC\,622$ which have not been
previously examined.
\end{itemize}
However, the problem is complicated by the fact that the general field which
contains all these newly catalogued quasars also contains another bright
($V=16.6\,mag$) active galaxy - the Seyfert $UM341$ ($z=0.399$) - which is also
a putative parent galaxy for any quasars in the vicinity.  What we shall
show is that {\it all} of the catalogued quasars within approximately 35 arcmins
of either of these two bright objects is a Karlsson peak object relative to one
or the other of them.

Table \ref{Table1} lists all catalogued quasars within 33' of $NGC622$ - with
the exception of two objects which we associate with $UM341$. Table 
2 lists all catalogued quasars which lie within 36' of $UM341$.
Fig. 4 shows a $40'$ radius field centered on $UM341$, with all
the $SDSS$ objects included. Note: Fig. 4 does not show all the
$NGC622$ objects, but does include two Karlsson-peak quasars (bottom $z=1.46,
3.63$) not listed in either table. These two are omitted because, although they
are Karlsson peak objects in the frame of $NGC622$ (or in the frame of either of
the other two $z=0.017$ objects in the field), they lie outside the chosen $33'$
radius circle associated with $NGC622$.
\begin{figure}
\includegraphics[  width=3.0in,
  height=2.8in]{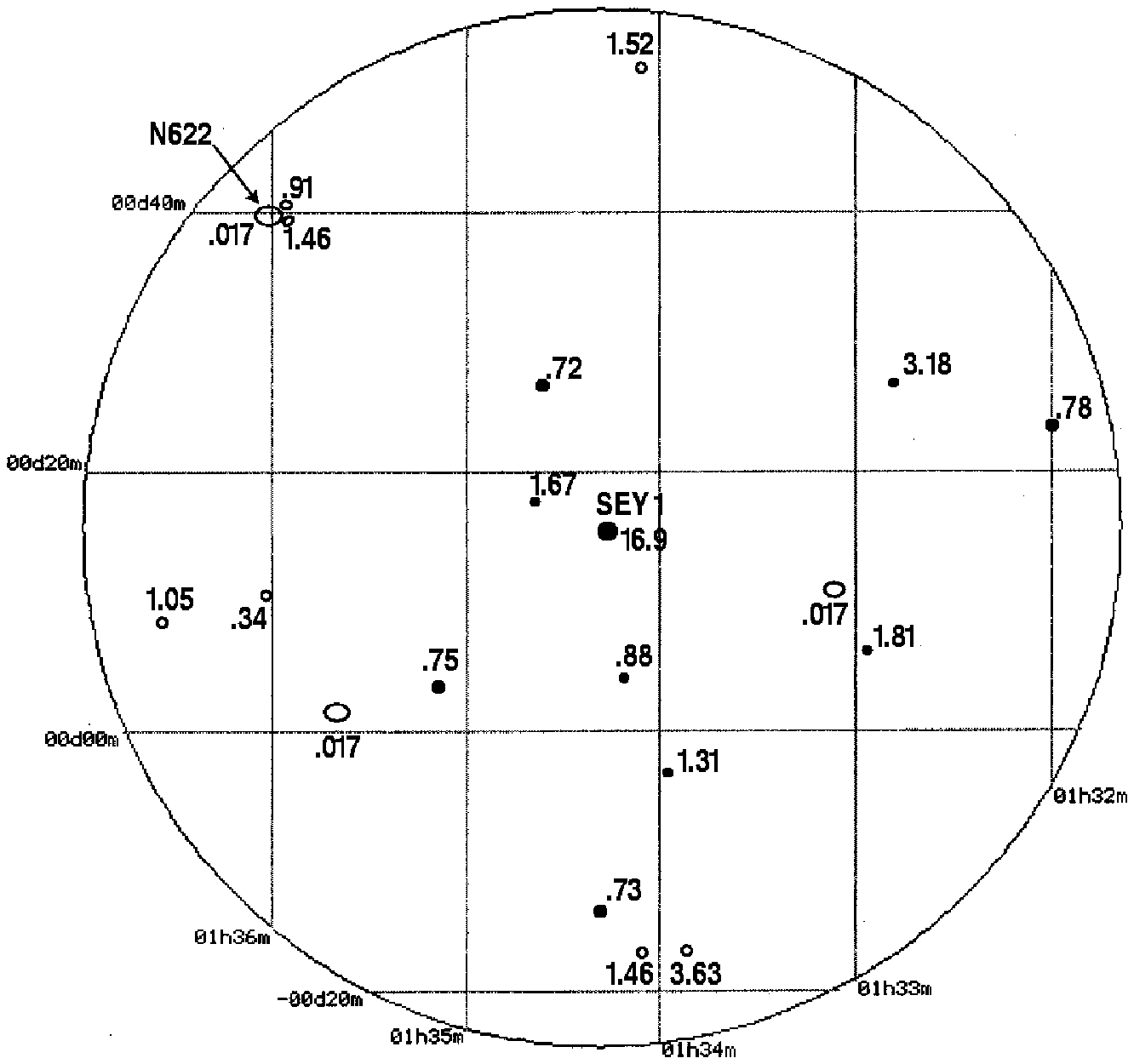}
 \caption{Simbad map showing all catalogued galaxies and quasars within $40'$
of $UM341$ (center). Open circles are Karlsson peak quasars in the frame
of $NGC \,622$ (or the other  $z = 0.017$ galaxies). Filled circles are Karlsson
peak quasars in the frame of the central bright Seyfert, $UM341$ ($z = 0.399$)}
\label{Fig4} \end{figure}

\begin{table} \caption{Quasars Associated with $NGC\,
622$} \label{Table1} \vspace{0.5cm}
\begin{tabular}{llrrrc}
Name & mag. (g) & $z$ & $z_0$ & $\Delta z$ peak &
Remarks\\
& & & & &\\
NGC 622 & m = 14.1 & .017& ---- & ---- & Mrk 571 parent\\
UB1 & 18.4 &0.910 & .88 & --.08 &\\
BS01 & 19.0 & 1.460 & 1.43 & +.02 &\\
SDSS & 18.9 & 1.501 & 1.46 & +.05 & \\
SDSS & 19.3 & 2.749 & 2.69 & +.05 & \\

FIRST& 17.8 & 0.344 & .32 & +.02 & Radio Gal\\
SDSS & 18.6 & 1.522 & 1.48 & +.07 & \\
SDSS & 19.2 & 1.049 & 1.01 & +.05 & \\
\end{tabular}
\end{table}

\begin{figure}
\includegraphics[  width=3.1in,
  height=2.8in]{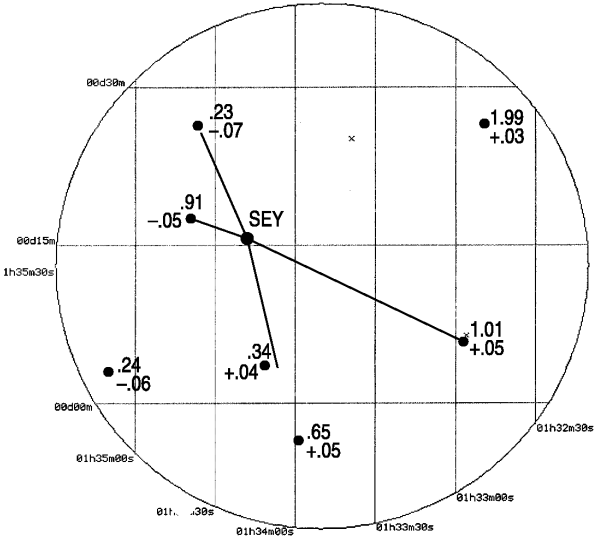}
 \caption{The Seyfert object (UM 341) and the seven nearest quasars associated 
with it. The quasar redshifts are transformed into the rest frame
of the central Seyfert object. The difference between this redshift ($z_0$) and the
nearest peak in the periodicity is written below the redshift.} \label{Fig4}
\end{figure}

\begin{table}
\caption{Quasars Associated with $UM \,341$}
\label{Table2}
\vspace{0.5cm}
\begin{tabular}{llrrrc}
Name & mag. (g) & $z$ & $z_0$ & $\Delta z$  peak &
Remarks\\
& & & & &\\
UM 341 & 16.6 & .399 & ----& ----& Seyfert parent\\
SDSS & 18.4 & 1.666 & .91 & --.05 & \\
SDSS & 18.6 & .718 & .23 & --.07 & \\
4C & V = 21.7 & .879 & .34 & +.04 & PKS B--V=.84\\
SDSS & 19.0 & .745 & .24 & --.06 & \\
UM 339 & 18.2 & 1.31 &.65 & +.05 &\\
SDSS & 19.3 & 1.805 & 1.01 &  +.05 & \\
SDSS & 21.9 & 3.183 & 1.99 & +.03 & \\
SDSS & 19.1 & .734 & .25 & --.05 & \\
SDSS & 19.1 & .781 & .27 & --.03 & \\
\end{tabular}    \label{Table2}
\end{table}

Tables \ref{Table1} and \ref{Table2} quantify the situation. Table \ref{Table1}
lists all the $NGC\,622$  objects, showing raw redshift data in the
$z$ column, corrected redshift data in the $z_0$ column and the difference
between $z_0$ and a Karlsson peak in the $\Delta z$ column.  Table \ref{Table2}
does the same for the bright Seyfert object, $UM341$.

In total, the two tables show {\it all sixteen} of the corrected quasar redshifts
lie, on average, within 0.049 of a Karlsson peak. When we realize that
the expected deviation from the low-value peaks ($z=0.3$) is about 0.075 and
from the highest value peaks ($z=2.64$) is about 0.17, and given that the
field was chosen because one of us (Arp) has already considered the same field
many years ago when it contained only two quasars, we can see that this is a
result having an {\it extremely} low probability  of being a chance result.

 {\it It is important to point out here
that we introduce in the appendix a new analysis which simultaneously measures the
goodness of fit of the redshifts to the periodicity law and the probability of finding the
redshift of the parent galaxy as measured.}

At his stage, however we already would like to emphasize that a partition of the 
sixteen quasars by their Karlsson peak properties corresponds closely to a partition 
according to their geometric association with each of $UM341$ and $NGC622$.  
Consequently if we had just done what was done in the highly publicized Hawkins
et al paper, then a negative result would most certainly have been obtained. 
The essential point is that Quasars must be correctly identified with their
putative parent galaxy, and moved into the rest-frame of that galaxy, if the
Karlsson peaks are to be observed.

\subsection{Geometric Centering of Putative Parent}

This type of disposition is observed in Fig. 4 where it is seen that the
Seyfert is roughly at the center of very similar redshift quasars, $z =
0.72$ to 0.78. In addition to this, the basic hypothesis leads us to expect 
(for reasons of momentum conservation) that quasar ejection events will tend to
occur as paired events with paired objects being ejected in opposite directions.

When looking just around UM 341 it is discovered that the inner four quasars form 
two rather well aligned pairs across the central Seyfert. They are marked in Fig. 5. 
It is seen that the deviations from the Karlsson peaks tend to be positive on one side 
of the pair and negative on the other. This is what would be expected if the pair
were ejected with a radial component of approaching velocity on one side and
receding on the other. Moreover the centering of the active galaxy is fairly good and 
{\it the quantitative values of the $\Delta z's$ are closely equal (e.g. +.05 and -.05 
across one pair).}

It is also to be noted in Fig. 5 that the rest of the quasars around UM 341 are
roughly aligned and the calculated ejection velocities are positive on the SW 
side and negative on the NE side.

Do such ejection velocities have any idependent precedent? Using the calculation      
                     $$1 + z_v = (1 + z_0)/(1 +z_{peak})$$ 
The actual current velocities compute to $\pm  7650\,km/sec$ and
$(+9230,\, -16150 )\,km/sec$ for the innermost two pairs. These velocities are
of the order of $10000\, km/sec$, similar to ejection velocities calculated
for most of the highly significant pairs ejected from well known Seyfert
galaxies (Arp 1998). Recent measurements of the initial outflow of the ionized material 
from the QSO/Seyert PG1211+143 give .08 to .10c (Pounds et al. 2003a,b). So there are 
direct confirmations of such velocities.

Many similar examples of ejection are presented in a new Catalogue of Discordant
Redshift Asssociations  (Arp 2003). 

\section[]{Visual Surveys in Three  $2dF$ and $SDSS$ Fields}

We should emphasize that the procedure we are following is to examine apparent groups and
concentrations of quasars that appear to be physically associated. In each of the cases
investigated here it turns out there is a brighter, active galaxy present which is a candidate for
the origin of these quasars. We then proceed to test this identification by seeing whether the
disparate redshifts are brought into the order of the Karlsson formula when their redshifts are
transformed to this chosen parent. 

We began by selecting three regions, chosen for reasons independent of the current 
considerations, from the $2dF$ and $SDSS$ surveys. Then, we simply performed an 
elementary visual inspection of quasar distributions within these fields, looking for the 
typical configuration signatures that the basic hypothesis leads us to expect.

\subsection{Region Selection Criteria}
The regions were selected by one of us (Fulton) from the $2dF$ and $SDSS$
surveys for an MSc study in data-mining {\it which was completely independent}
of the present analysis (Fulton 2002).  For illustration of the data-mining
technique, it was decided that a comparative study of galaxy distributions and
quasar distributions over the same areas of sky would be ideal. Thus, a
requirement for reasonably complete surveys of both galaxies and quasars in the
chosen areas was generated.  Since the Fulton study was using early
releases of $2dF$ data, this imposed particularly tight constraints on the
choice of suitable $2dF$ fields.

Fulton was able to locate one suitable region from $SDSS$ and two from $2dF$.
These regions are:
\begin{itemize}
\item $2dF$ region 1:  Centered on $RA(23h24m30s)$ and $Dec( -28d33m36s)$ and
covering about $6$ square degrees;
\item $SDSS$ region 2: Centered on  $RA(09h50m00s)$ and $Dec(00d00m00s)$ and
covering about $10$ square degrees;
\item $2dF$ region 3: Centered on  $RA(13h41m)$ and
$Dec(-01d30m00s)$  and also covering about $6$ square degrees.
\end{itemize}

\subsection{Region 1 - A $2dF$ Field.}
The basic hypothesis leads us to expect the existence of strings/groups of
quasars associated with bright active galaxies.  A visual examination of
Fulton's region 1 leads to a straightforward identification of such an apparent
grouping within  a $35'$ circle, centered on $RA(23h11m30s)$ and
$Dec( -28d15m00s)$.  Simbad was used to get a complete listing of all the other
objects in this field, which is shown with all its catalogued contents in
Fig 6.   It is clear that the quasars in the field form a  rather
dense cone-like distribution having at its apex the very bright and
very low redshift object  $NGC \,7507$ and the companion galaxies that make up
its group (Arp/Madore 2309-284).  If  $NGC\,7507$ is
taken as the putative parent then, according to the basic hypothesis and since
this object has an extremely small redshift at $z=0.005$, we should expect to
find a strong bias to the Karlsson peaks in the raw data.

In Figure 6 the observed redshifts are written next to each quasar. Filled
circles indicate redshifts falling within the $50\%$ probability circle around the 
Karlsson peaks (that is, within a region $\pm0.022$ centred on each peak on the
$\log_{10}(1+z)$ axis), and open circles indicate redshifts falling outside the
$50\%$ circle. It is clear that the quasars stretching up and to the NW from
$NGC \,7507$ are almost all near the Karlsson peak values.
\begin{figure}
\includegraphics[  width=3.2in,
  height=2.8in]{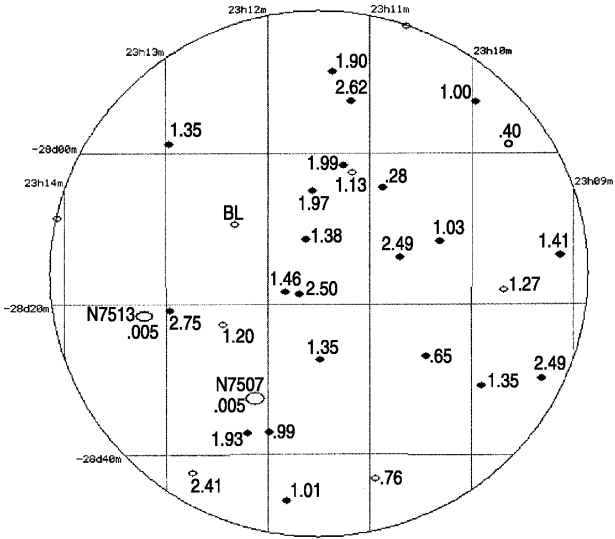}
\caption{2dF quasars inside a $35'$ circle centered at 23h11m30s
-28h16m00s. Redshifts near the Karlsson peaks are indicated by filled circles,
redshifts between peaks  are indicated by open circles. NGC 7507 is the
principle galaxy in a group with redshift $z = 0.005$.} \label{Fig5}
\end{figure}
\begin{figure}
\includegraphics[  width=3.1in,
  height=2.0in]{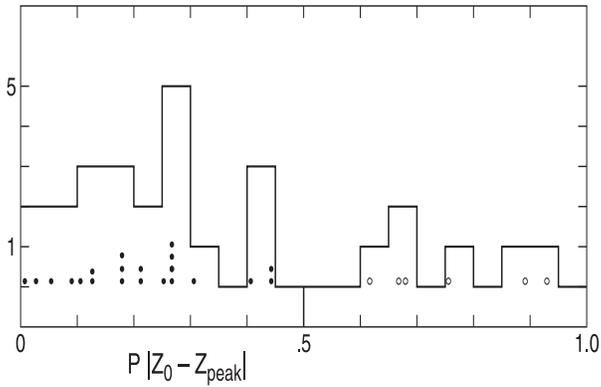}
 \caption{The distribution of probabilities of the redshifts falling close to the peaks for 
all the quasars in Fig \ref{Fig5}. Open circles illustrate redshifts falling away from
the peaks, filled circles the 21 quasars falling close to the peaks.}
\label{Fig6} \end{figure}

The actual distribution of these redshifts relative to the Karlsson peaks is
shown in Figure 7. The numerical probability of this observed distribution 
being due to chance is discussed in Appendices A and B.

But in addition to the high probabilities for the realities of the periodicity, 
the physical association of these quasars with the bright galaxy is 
reinforced because it is possible to draw a cone shaped perimeter, with NGC 7507 
at one end, which contains 15 quasars, {\it only one of which does not fall near a 
Karlsson  peak}. It seems that this would strongly support the ejection origin for the 
excess number of quasars from this very low redshift (z = .005), bright galaxy.

There are two quasars close to the SSE of NGC 7507 in Fig. 6 but further to the SE 
there is a paucity of 2dF quasars as if we were seeing background values. There are 
some active galaxies which appear to have one sided ejections, for perhaps the same
reasons that there are one sided jets from galaxies. One could suggest that ejections
in certain directions are deviated or broken up upon exiting through certain  portions
of the parent galaxy.

\subsection[]{Region 2 - An $SDSS$ Field}
The basic hypothesis requires that Karlsson peak quasars are associated with
{\it specific} putative parent active galaxies that, generally speaking,
will be much brighter than their offspring quasars. The general ejection picture
suggests that, commonly, we should find quasars with matching redshifts paired
across their putative parents, and many such configurations are already known.
(Arp 1967; E.M. Burbidge 1995; Arp et al. 2001). By contrast, if we
accept the canonical viewpoint about quasars, then such configurations should be
very rare.

It is notable, therefore, that one such configuration is readily identifiable in
Fulton's Region 2:  We consider a $30'$ circle, centered on $RA(09h49m48s)$ and
$Dec( 00d37m30s)$ from this region.
Fig. 8 shows six quasars in a field of radius $30'$ which are
centered on the active triplet of low-redshift galaxies around $NGC \,3023$
($z=0.006$). {\it The outstanding pair consists of $z = 0.640$ and $z = 0.584$
quasars which fall $z_v = +0.02$ and $- 0.02$ from the major redshift peak at $z
= 0.60$.} It would be difficult to avoid the implication that they had been
ejected from one of the central galaxies and were now travelling with a radial
component of velocity $0.02c$, one away from, and one toward the observer.
\begin{figure}
\includegraphics[  width=3.1in,
  height=2.8in]{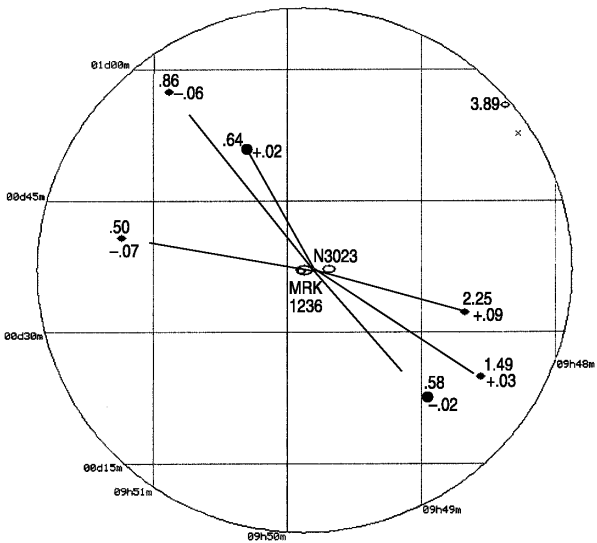}
 \caption{All SDSS quasars falling within $30'$  of the $NGC\, 3023$
($z=0.006$) group (radio galaxies plus $Mrk\, 1236$ ($z=0.006$)). The
principle pair is at $z = 0.64$ and $z =0.584$ (both double radio sources).
Redshifts and $z_v$ differences from Karlsson peaks are written next to quasar
symbols} \label{Fig7} \end{figure}

Moreover, there are two other pairs of quasars in approximately the same direction. 
One pair has apparent velocity deviations from the redshift peaks of $z_v=+0.09$ and
$-0.07$ and the other  $z_v=+.03$ and $-0.06$. This pattern is
characteristically encountered (e.g. see $UM\, 341$ in the $NGC\, 622$ field and
pairs analyzed in the introduction to "A catalogue of Discordant Redshift
Associations" (Apeiron 2003). The chances would seem vanishingly small to
find repetitions of such patterns in random associations of background objects.
In the present case, however, there is even more evidence for association in the
fact that both members of the major pair at $z_{peak} = 0.60$ are strong radio sources.
The central galaxies are both NVSS radio sources and the {\it the two quasars
are each double radio sources}. The latter is quite unusual and represents
additional evidence against accidental association.

It should be noted that the central galaxies here are low redshift so that only small 
corrections to their rest frames are needed.

\subsection[]{Region 3 - A $2dF$ Field}
Finally, the software was used in an experiment to see whether we could find an 
example of a group of quasars all at the same Karlsson peak. 
The point was to see if there was a bright, active galaxy assignable to the origin 
of the group. We found the string of z = 2.64 ($\pm0.1$) quasars pictured in Fig. 9.  
We were pleased to find a bright Seyfert/QSO at the center of this string.

But unfortunately (or so it seemed),  the bright Seyfert/QSO
had a large redshift of $z=0.236$, which seemed certain to destroy the Karlsson
peak coincidence, after the rest-frame transformation. However, when the six
quasar redshifts were transformed into  the rest-frame of $UM602$, they became 
$z = 1.92, 1.99, 2.02, 1.86, 2.01,1.86$ (mean value $z=1.94$). These are
all clustered around the major Karlsson peak at $z=1.96$ (first discovered at 1.95). 
Moreover the innermost quasars form rather well aligned pairs across UM 602. 
One pair is z = 2.72 and z = 2.73, the other pair is z = 2.54 and z = 2.54.

\begin{figure} \includegraphics[ width=3.1in, height=2.8in]{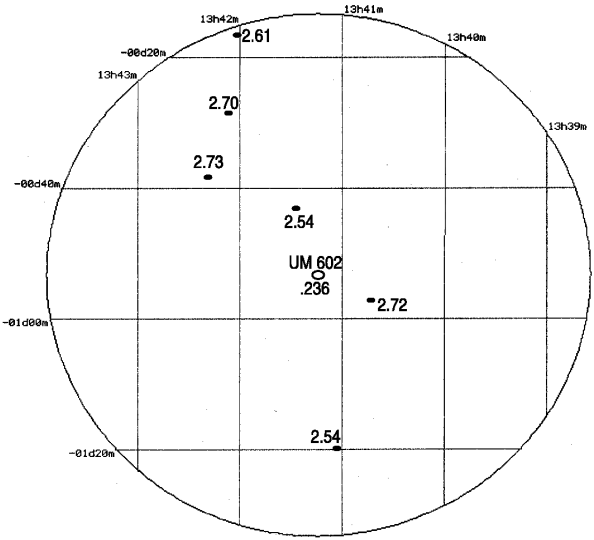}
\caption{$UM \,602$, a bright QSO/Seyfert object is the center of six quasars
which are close to the $z = 2.64$ Karlsson peak. When transformed to the $z =
0.236$ redshift of $UM\, 602$ the mean redshift of the six comes out $z = 1.94$,
close to the Karlsson peak of $z = 1.96$.} \label{Fig8} \end{figure}

\section[]{Objects With Two Redshifts}
In a discussion of spectra in the 2dF fields, Madgwick et al. 2002 reveal the occurrence 
of a low redshift galaxy of z = .16 and a quasar {\it in the same spectrum} of z =.87! They 
note it is "a spectrum showing evidence for a low redshift galaxy and a quasar at much
higher redshift" but with no evidence for a gravitational lens. It is  
interesting to note that if the quasar is physically associated with the galaxy its redshift
would be transformed into z = .61, very near the Karlsson peak of z = .60. This is also
support for the result  of Burbidge and Napier 2001 who tested, with positive results,
redshift periodicity in apparent pairs of quasars separated by $\leq 10^".$

Another case of two redshifts in one spectrum is 3C343.1, a radio galaxy with a bridge to a
quasar which is only .25 arcsec distant. The galaxy has z = .344 and the quasar z = .750.
When transformed to the z = .344 galaxy, however, the quasar redshift becomes $z_0 = .302$
which is rather close to the Karlsson value of $z_{peak} = .30$ (Arp, Burbidge and Burbidge ,
2004.)

\section[]{Summary}
For over 35 years now the evidence has been building for a set of numerically defined 
peaks in
the distribution of quasar redshifts - the so-called Karlsson peaks.  But the
existence of the Karlsson peaks has generally not been acknowledged
 - primarily because of the serious implications for canonical cosmology.

The new large scale $2dF$ and $SDSS$ surveys offer the opportunity to settle the
issue finally using quantitative statistical methods - but, because of the
fundamental importance of the claims, it is crucial that such large
scale studies are rigorously designed and executed. 

We emphasize (a) that evidence supporting the basic hypothesis is readily found
in the new surveys and (b) what is required is a rigorous analysis of the areas 
investigated.  We have made a variety of tests in sample areas of these surveys and 
have found that periodicity of redshifts is strongly present in contiguous groups
of quasars - but only in the reference frame of the associated, dominant galaxy. This 
result not only strengthens the  universality of the periodicity relation but also 
confirms the physical association of higher redshift quasars with relatively low 
redshift, generally brighter galaxies. The parent galaxies further confirm the
physical associations by turning out to be generally emission lines objects,
often morphologically disturbed and often showing X-ray and radio emission

\appendix

\section{Analysis of Periodicity by a  Method of Minimum Residuals}

\begin{figure} \includegraphics[ width=3.1in, height=2.0in]{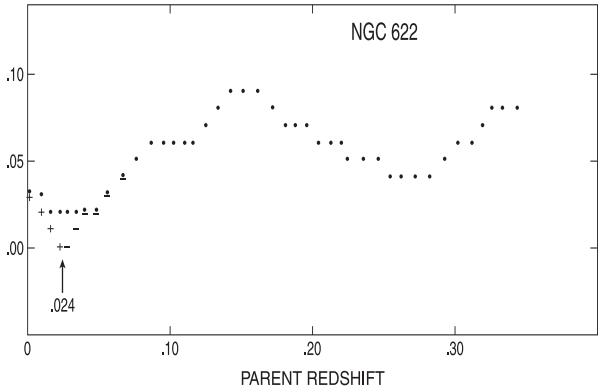}
\caption{NGC 622 is the  bright Seyfert at the center of seven quasars in Fig. 3.
When transformed to the rest frame of galaxies having a range of redshifts, the 
residuals from the nearest Karlsson peaks (measured in $z _v's$) assumes a deep 
minimum at z = .024. The actual redshift of NGC 622 being z = .017 indicates the 
association is unlikely to be chance and that standard periodicity is accurately 
fitted. Absolute values of the residuals are plotted as dots.} \label{Fig10} \end{figure}

     The question arises as to whether it is possible to test quasars in a group around 
a candidate parent galaxy against control fields or by monte carlo methods. It turns out 
there is, by the simple expedient of transforming redshifts to the rest frame of the 
presumed parent galaxy and then varying the redshift of that parent. We have done that 
here by transforming  the quasar redshifts listed for NGC 622 and UM 341 in Tables 1 
and 2 to $z_o$ by means of the redshift of the central galaxy, then computing $z_v$, 
the component of velocity needed to explain its deviation from the nearest, exact 
redshift peak. We then compute the same $z_v$'s after transforming the observed 
redshifts to a large range of different parent redshifts.

\begin{figure*} 
\includegraphics[ width=6.2in, height=2.5in]{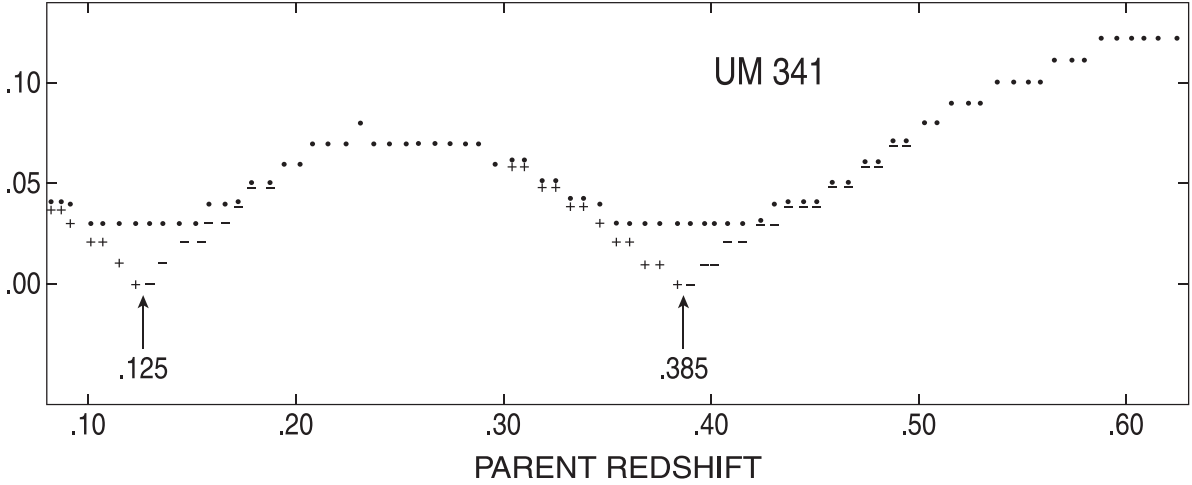}
\caption{UM 341 is the Seyfert at the center of nine quasars in Fig. 4.
When transformed to the rest frame of galaxies having a range of redshifts, the 
residuals from the nearest Karlsson peaks (measured in $z _v's$) assumes a deep 
minimum at z = .385. The actual redshift of UM 341 being z = .399 indicates the 
association is unlikely to be chance and that standard periodicity is fitted with great
accuracy. Another minimum almost as well defined, at z = .125, is almost exactly one 
Karlsson period away.} \label{Fig11} \end{figure*}

The results are shown in Figs. 10 and 11. Plotted are the mean of the absolute values
of the residuals, $<|z_v|>$, from the periodicity formula (plotted as dots). Plotted also are 
the mean values of the residuals, $<z_v>$, with plus for a positive and minus for a negative 
residual.  What we see in Fig. 10 is that the scatter for the  absolute values of the residuals 
remains large throughout the tested range. But at a low redshift, about that of the parent, 
z = .02 - .03, the scatter reduces to about .03 to .04 and the residuals then start a smooth 
convergence, sytematically positive, reducing toward zero and then increasing negatively 
on the other side of a sharply defined position where the receding and approaching 
ejection velocities are at a minimum and exactly balanced between plus and minus.

That value in Fig. 10 is about z = .024. This is to be compared to the actual value of the
observed parent galaxy of z = .017.  This is the  only value in the tested range which gives 
a very close fit to the Karlsson formula series of redshift peaks. It seems that we have
shown that this group of disparate redshifts shows almost perfect correspondence to the
periodicity peaks but only when transformed to a redshift frame very close to the
previously assigned parent redshift.
 
Fig 11 shows a similar analysis applied to the nine quasars around UM 341. Here we see 
two places where the absolute values of the residuals becomes very small and the the 
plus and minus values balance at zero. One is z = .385, close to z = .399, the observed 
value for UM 341. The other is z = .125, almost as good a minimum, and interestingly, 
(1 + .385)/(1.23) = (1 + .126),  an almost exact Karlsson interval distant. The latter suggests 
that the Karlsson period remains exact even when the parent galaxy falls somewhat 
inexactly on a peak.

A word should be mentioned about the plus and minus signs between minima. When the
transformed z's transit from one peak to the next their residual is at a maximum but
suddenly changes sign, affecting the mean strongly. It is not until the residuals are all small,
that the plus and minus $z_v$'s accurately measure the convergence of balanced plus and
minus ejection velocities. The absolute value of the residuals has a shallow minimum no
more than $<|z|>$=.03 suggesting that it is the average projected ejection velocity which then
must average to near zero in pairs of oppositely ejected quasars. That average projected
velocity is comparable with the measured outflow velocity of ionized material close to AGN 
quasars, e.g. z = .08 to .10 (Pounds et al. 2003).  

When the minimum residual analysis is applied to the 27 quasars in the field of NGC 7507 
(residual plot not shown here) the minimum is clearly at z = .005, the measured redshift of the 
parent galaxy. When it is applied to just the 15 quasars in the cone going NW, however, the 
minimum at z = .005 becomes sharper and more conspicuous. In the probability analyses of 
Appendix B to follow, a power spectrum analysis of the 27 quasars in this 2dF field is shown.

\subsection{A group of quasars south of NGC 2639}

Finally we should mention that this same analysis can be applied to a group of 10
quasars south of the bright Seyfert, NGC 2639, discovered in the era when quasar detection 
was done by ultraviolet excess (Arp 1980). In this case the minimum residual analysis is now 
especially accurate in picking out the actual redshift of the real parent and confirming the 
behavior of the residuals. In Fig. 12 we show the distribution and redshifts of the QSOs 
around the companion galaxy which has z = 0.006. Arp originally argued that these QSOs 
originated in that companion. However, a spectrum of this galaxy shows only absorption 
lines and thus no sign of nuclear activity (Arp 1980).

\begin{figure}
\includegraphics[width=3.1in] {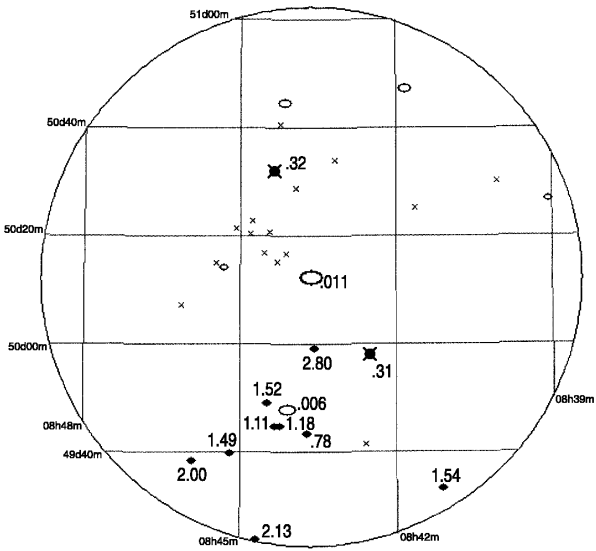}
\caption{A 50 arcmin radius around NGC 2639 showing known X-ray
and optical objects (from SIMBAD). Cataloged redshifts are
labeled next to the known quasars in the group around the
southern companion galaxy at z = .006.} \label{fig12}
\end{figure}

We consider instead the possibility that these QSOs have
originated from the brightest active system in their vicinity
which is the QSO with z = 0.305.  We have tested this hypothesis
by transforming the observed redshifts into the two reference
frames, dividing by 1.006 and then by 1.305.  The results are
given in Table A1.

\begin{table} \caption{Quasars South of $NGC\,
2639$} \label{Table3} \vspace{0.4cm}
\begin{tabular}{llrrrc}
Quasar & ~$z$ & $z_o (0.006)$ & $z_v$ & ~~~~~~~~$z_o (0.305)$ & $z_v$\\
& & &\\
U 10 &  0.305 &   0.297 &   0.000  &   ---- & ----\\
U  8 &  2.800 &   2.780 & + 0.040  &  1.912 & - 0.02\\
U  3 &  1.522 &   1.507 & + 0.040  &  0.933 & - 0.01\\
U  1 &  1.177 &   1.164 & + 0.100  &  0.668 & + 0.04\\
U  2 &  1.105 &   1.092 & + 0.070  &  0.613 & + 0.01\\
U  4 &  0.780 &   0.769 & + 0.106  &  0.364 & + 0.05\\
     &        &         & - 0.097  &        &       \\
U  5 &  1.494 &   1.479 & + 0.030  &  0.911 & - 0.02\\
U  7 &  2.000 &   1.982 & + 0.010  &  1.300 & - 0.05\\
U 14 &  2.132 &   2.114 & + 0.050  &  1.400 &   0.00\\
U 15 &  1.535 &   1.520 & + 0.050  &  0.943 & - 0.01\\
\end{tabular}
{\em

~~~~~~~~~~~~~~~~~~ $<z_v>$ = +.03 to +.05~~~~~~~~~~$<z_v>$ = .00

~~~~~~~~~~~~~~~~~~~~~~$<|z_v|>$ = .049~~~~~~~~~~~~~~~$<|z_v|>$ =
.023\/}
\end{table}

We can then see which set of transformed redshifts fits the
periodicity peaks best. The answer is given at the bottom of the
4th and 6th columns. It shows that the relative velocities
measured by $z_v$ for the companion case are all positive. (Except
for U4 which sits just on the edge between two Karlsson peaks). On
the other hand the plus and minus values exactly balance for the z
= 0.305 case. This means the average ejection velocities in the
approaching and receding directions average to zero as they should
if they are associated with this QSO. (The same argument holds if the 
residuals are not velocities but represent intrinsic scatter around a peak.) 
Also the absolute size of the residuals needed to correct the redshifts on 
to the intrinsic peaks are only half as large than when it is assumed that the
z = 0.006 galaxy is the progenitor.

In order to further illustrate the significance of the periodicity present in
the redshifts of the group of QSOs south of NGC 2639, we now plot the 
residuals from those redshifts in the rest frame of a range of
possible parent galaxies. {\it Fig. 13 shows that the deviations
from the Karlsson periodicities reach a very sharp minimum exactly
at z = .305, the redshift of the actual AGN/QSO that we have
assigned as the parent of this group.}

\begin{figure*}
\includegraphics[width=6.2in, height=2.5in]{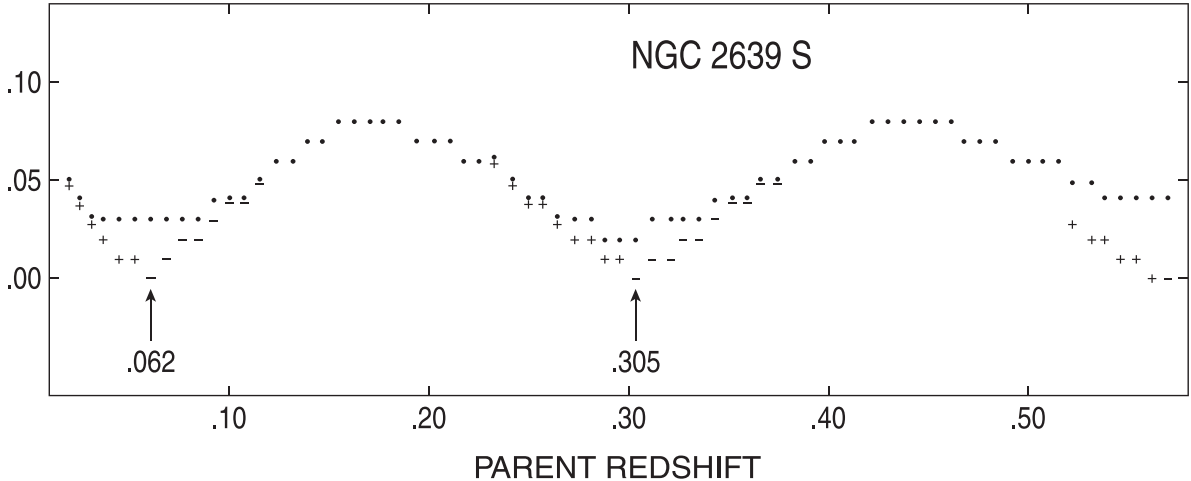}
\caption{The AGN with  z = .305 (southern of the pair across NGC
2639) is taken as the ejecting parent of the nine quasars in Fig. 12. When 
transformed to the rest frame of galaxies having a wide
range of fictitious redshifts, the  residuals from the nearest
Karlsson peaks (measured in $z _v's$) assumes a deep minimum just
at z = .305. Therefore the association is extremely unlikely to be
chance. The standard periodicity is thus fitted with great
accuracy by this group of quasars discovered in 1980. A secondary
minimum at z = .062 is exactly at the next  Karlsson peak lower. The residuals are
given near the minima as + and - symbols. The absolute values of the residuals
are given as dots.}
\label{fig13}\end{figure*}

It is also apparent that there is a secondary minimum for a parent
at z = 0.062, not quite as good as the z = 0.305 but very similar
and exactly one Karlsson period away from z = 0.305.  Early
studies of low redshift QSOs and AGN suggested that the first
intrinsic redshift peak lies at z = 0.061 (Burbidge 1968).  In a
later sample Arp et al. (1990) obtained z = 0.062.

A similar situation suggesting secondary ejection might also be
true for the northern QSO with z = 0.323 which is paired with the
z = 0.305 QSO across NGC 2639.  However, to test this it would
first be necessary to search for QSOs around this object, and so
far this has not been done.

Here we must gratefully acknowledge the remeasuring of six of the quasars in the 
NGC 2639S group which resulted in correction of two of the original 10 redshifts
(Ford, H. et al. 1983). These corrections helped make the fit of the quasars to the 
periodicity, and their association with the parent, so precise in the present paper,

In all, this analytical procedure raises interesting opportunities to study the mathematical 
behavior of the Karlssson periodicities as manifested by parents of different redshifts. 
Taking a group of quasars of apparently unrelated redshifts, making the required
transformation to the redshift of the candidate parent galaxy and then finding the redshifts 
to fit a previously well defined formula, would seem to confirm at a very high level of 
probability the periodicity relation as well as demonstrating again the physical 
association of specific low redshift galaxies with high redshift quasars.

\section{Numerical Estimates of Probability}

In the fields around NGC 622 and UM 341 we note that on the $\log_{10}(1+z)$ axis the 
Karlsson period is $0.089$, so that the expected deviation from a peak-value is about 
$0.022$. The mean actual deviations of the $\Delta z$'s is $0.012$, and a very 
conservative calculation of the odds of all fourteen of the {\it new} quasars being a 
Karlsson peak quasar relative to either one of $NGC622$ or $UM341$  gives a probability 
of $2.6\times 10^{-5}$ of it being a chance occurrence. 

In more detail, we ask the following question: given the existence of the two bright
active independent objects, $NGC622$ and $UM341$ in the field of figure
Fig. 4 and the two existing Karlsson peak quasars (Arp 1981, 1987) adjacent
to $NGC622$, what is the probability of fourteen out of fourteen  new quasars
turning out to be Karlsson peak objects for {\it either one} of $NGC622$ or
$UM341$?

We proceed as follows: the expected dispersion of measured redshifts from the
nearest Karlsson peak along the $\log_{10}(1+z)$ axis, given that there is no
effect, is $0.022$. In fact, the actual mean dispersion  is $0.012$, and the
odds of any one quasar falling that close to a Karlsson peak after being
transformed into the frame of, say $NGC622$, is about $0.27$ - given that there
is no effect.

But, the quasar might be a Karlsson peak object for {\it either} of $NGC622$ or
$UM341$, and since we allow ourselves this possibility, it must be accounted
for. The probability of this dual eventuality is easily calculated to be
about $0.47$. Since we have fourteen such quasars, the overall probability is
now estimated as $0.47^{14}=2.6\times10^{-5}$.

This calculation is highly simplified of course, but it is to be noted that no
account is taken of the pairing configuration of Figure \ref{Fig4} - even though
such configurations are to be expected in terms of the basic hypothesis and not to 
be expected if the quasars are unassociated. Thus the true probabilities of finding 
what is observed are very much smaller than we actually calculate.

As for the results shown in Figs. 6 and 7, one approach would be to reason 
that if the redshifts were not related to the specific periodicity peaks one would expect 
equal probability of the points in Fig. 7 falling anywhere in the interval P = 0 to 1 with a 
mean at P =.5. - i.e. a horizontal line. The histogram shows, however, an excess of 14
points with P = .28 and less.The chance of this can be roughly calculated as 
$P(.28)^{14}/P(.5)^{14} = 3\times10^{-4}$. 

Alternatively we could estimate the probability of the
configuration arising by chance, by noting that of
the 27 quasars in the field, 17 of them actually fall within the $28\%$ circle
around the Karlsson peaks and 10 fall outside. Simple binomial statistics then
gives the probability of this configuration arising by chance, as 
$1.5\times10^{-4}$. 

Finally Fig. 14 gives a power spectrum analysis of the 27 quasars in the region 
around NGC 7507 as shown in Fig. 6. The period is P = .083 and the power I = 15.0.
I 

\begin{figure}[h]
\includegraphics[  width=3.5in,   height=3.5in]{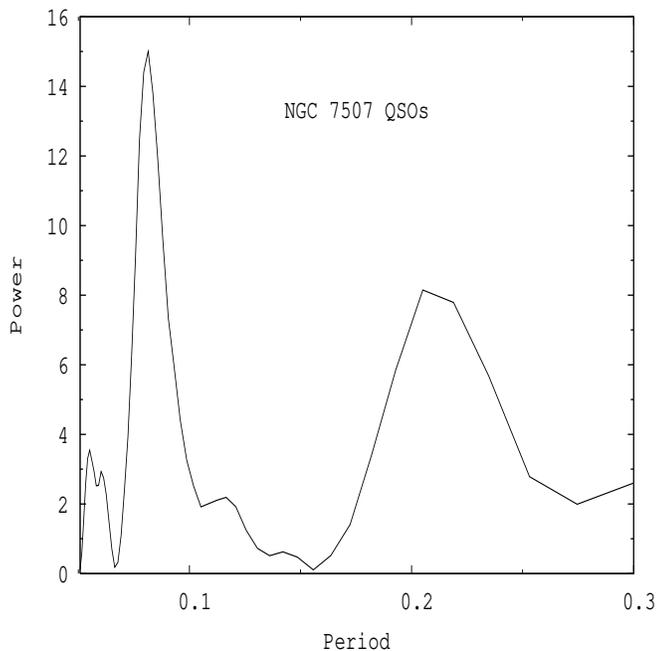}
 \caption{Power spectrum analysis of quasars in the NGC 7507 region which is 
pictured in Fig. 6. Power is around I =15 and period is close to Karlsson period 
(see text).}
\label{Fig14}  \end{figure}

If we then test just the quasars in the NE cone emanating from NGC 7507, the 
period becomes P = .086 and I = 15.6. The canonical Karlsson value for the 
periodicity is ${\Delta log(1 + z) = .089}$. 

All of these various tests appear to give a significant result in agreement with 
the visual impression.

\end{document}